# Thermoelectric power factor limit of a 1D nanowire

*I-Ju Chen, Adam Burke, Artis Svilans, Heiner Linke, Claes Thelander*

Solid State Physics and NanoLund, Lund University, Box 118, 22100 Lund, Sweden

Abstract

In the past decade, there has been significant interest in the potentially advantageous thermoelectric properties of one-dimensional (1D) nanowires, but it has been challenging to find high thermoelectric power factors based on 1D effect in practice. Here we point out that there is an upper limit to the thermoelectric power factor of non-ballistic 1D nanowires, as a consequence of the recently established quantum bound of thermoelectric power output. We experimentally test this limit in quasi-ballistic InAs nanowires by extracting the maximum power factor of the first 1D subband through *I-V* characterization, finding that the measured maximum power factors conform to the theoretical limit. The established limit predicts that a competitive power factor, on the order of mW/m-$K^2$, can be achieved by a single 1D electronic channel in state-of-the-art semiconductor nanowires with small cross-section and high crystal quality.

Thermoelectric devices can convert heat gradients into electricity, or pump heat by using electricity, and thus have applications both in energy harvesting and solid-state refrigeration. Long-standing challenges for a widespread use of thermoelectric applications are to find material systems with increased energy conversion efficiency as well as higher power output. For a temperature difference $\Delta T \ll T$, the maximum power output is proportional to the power factor of the material, $S^2\sigma$, with Seebeck coefficient $S$ and electrical conductivity $\sigma$. Hicks and Dresselhaus' pioneering theoretical work [1] identified that high $S^2\sigma$ can be achieved by one-dimensional (1D) charge channels in extremely thin quantum wires with high mobility. A similar analysis, also based on the Boltzmann transport equation, approximated the electron scattering time by a power-law of the energy $E$, $\tau(E) \sim E^r$, and indicated that $S^2\sigma$ can increase, in principle, indefinitely with increasing mobility, scattering parameter $r$, and decreasing nanowire cross-section area. [2] In this picture, a NW can potentially have an unlimited power production ability. The expected large $S^2\sigma$ in nanowires (NWs), together with their low thermal conductivity, led to the prediction of high energy conversion efficiency, triggering widespread efforts to develop NW-based thermoelectric materials.

Experimentally, however, the predicted high $S^2\sigma$ based on 1D electronic transport in NWs has to date not been observed. Thermoelectric properties of ballistic quasi-1D systems, including conductance quantization at multiples of $2e^2/h$ and oscillating Seebeck coefficient as a function of gate voltage, were first studied in quantum point contacts (QPCs). [3,4] At that time, the values of power production or power factors were not explicitly extracted. In NWs, 1D effects are often obscured due to scattering and formation of quantum-dot-like states [5], nevertheless, conductance plateaus and Seebeck coefficient oscillations have been observed [6,7]. However, accurate extraction of reliable values for the power factor in 1D NWs remains difficult for a number of reasons: first, as pointed out in Refs. 6 and 7, it is difficult to measure the Seebeck voltage at high NW impedance, the condition under which the peak power factor is expected; second, because the transport properties of single-NW devices are susceptible to the device length and defect distribution, it is critical to measure the conductance and Seebeck voltage simultaneously on the same NW segment, which can be challenging.

Recently, Whitney established that the power production by a 1D channel is intrinsically limited by quantum effects. [8,9] This prediction, which is based on nonlinear scattering theory, implies that a 1D electronic channel only has a limited ability to produce power through the thermoelectric effect, described by its power factor $S^2G$ with conductance $G$. It is worth noting that the term power factor is used for both $S^2\sigma$ and $S^2G$, which are related to the thermoelectric power density and power, respectively. $S^2G$ is often preferred for mesoscopic systems where a local description of electrical conduction is not adequate [10]. Nonetheless, the quantum limit is only attainable when the transport is ballistic. Therefore the question arises: What is the maximum $S^2G$ that can be measured directly in non-ballistic 1D electronic channels, and as a result, what is the achievable $S^2\sigma$ in realistic NWs?

Here, we address this question by establishing a theoretical upper bound for the power factor (both $S^2\sigma$ and $S^2G$) of a non-ballistic 1D NW, and by testing this limit through measurement of the power factor of single-InAs nanowire devices. First, based on the theory of the quantum bound of thermoelectric power output and considering non-unitary electron transmission, we formulate a theoretical limit of $S^2G$ in non-ballistic 1D channels. Then, in the experiment, we study conductance quantization and Seebeck coefficient oscillations, which are the characteristics of 1D subband transport. We use current voltage (*I-V*) characterization to directly and simultaneously measure the electrical conductance and Seebeck voltage on the same InAs nanowire segments. We demonstrate that the theoretical limit is consistent with the measured $S^2G$ maximum. Finally, by considering that the transmission probability of electrons scale classically with device lengths, we establish the limit of $S^2\sigma$ in non-ballistic 1D NWs to provide an indicator for the optimal $S^2\sigma$ that can be achieved with 1D charge transport in realistic NW structures.

For ballistic 1D channels, Whitney derived that the maximum power output is equal to the quantum bound $P_{QB} = B_0 \frac{\pi^2}{h} k_B^2 \Delta T^2$ with $B_0 \approx 0.0321$. [8,9] Thus, the power factor quantum bound for a spin degenerate 1D channel is

$$(S^2G)_{QB} = 2 \times 4B_0 \frac{\pi^2}{h} k_B^2 \approx 0.73 \ (\text{pW/K}^2), [11] \tag{1}$$

based on the relation $P_{max} = (S^2G) \times \Delta T^2/4$ [2], where $P_{max}$ is a thermoelectric system's maximum power output, and the factor of 2 is due to spin degeneracy. In quasi-ballistic and diffusive 1D channels, the transmission probability of charge carriers is less than unity, and we can extend the derivation of $(S^2G)_{QB}$ to provide a power factor upper bound also for these transport regimes. We first note that a step function is the optimal transmission function shape to maximize thermoelectric power output, and second, that the power output increases linearly with the step height, as pointed out in refs. [8,9]. Thus, we obtain the theoretical limit of the power factor of non-ballistic 1D channels

$$(S^2G)_{limit} = T_{max} \times (S^2G)_{QB} \tag{2}$$

where $T_{max}$ is the maximum of the energy dependent transmission probability of the 1D electron channel.

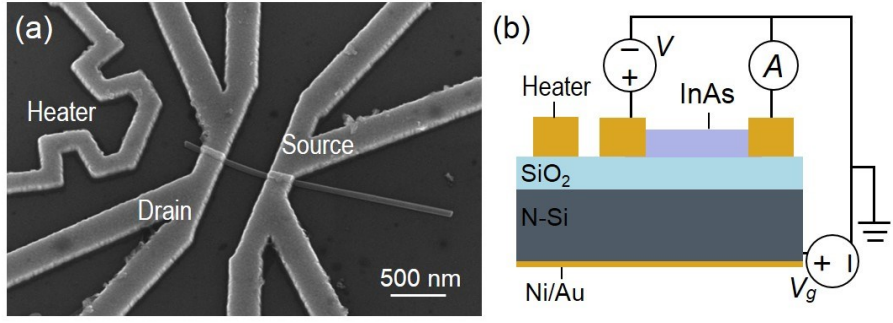

**Figure 1.** (a) Scanning electron micrograph of the InAs NW back-gate field-effect transistor with a side Joule heater. The experimental data of this exact device is displayed in Fig. 3(a, d). (b) Schematic of the device and circuitry, where a source-drain bias $V$ and a back-gate voltage $V_g$ are applied to the device.

In order to test this limit, InAs (zinc blende) NWs with 60 ± 4 nm diameter were grown by metal-organic vapor phase epitaxy (MOVPE) [12] and used to fabricate single-NW back-gate field-effect transistors with side Joule heaters. NWs are first deposited onto $SiO_2$ (150 nm)/Si substrates, where the degenerately $n$-doped (P) Si acts as a back-gate. Then the NWs are imaged by low-resolution scanning electron microscopy and selected for contact processing. The samples are spin coated with resist (polymethyl methacrylate, PMMA) and electron beam lithography is used to create openings for the NW contacts and the heater circuit. The InAs NW contact areas are ashed with $O_2$ plasma and passivated in a mixture of $(NH_4)_2S_x$ and $H_2O$ 1:20 for 1 minute at 40°C before Ni/Au contacts are evaporated onto the sample and then lifted off in acetone. An example of a finished device is shown in Fig. 1(a). We measure the conductance and Seebeck voltage with an electrical circuitry shown in Fig. 1(b), and $\Delta T$ with resistance thermometry using the source and drain 4-probe metal lines (supplemental material) in a variable temperature probe station. In some cases, $\Delta T$ is measured in the absence of the NW or on a different device with approximately identical structure. However, negligible variations are expected in the measured $\Delta T$ as the temperature profile along the substrate surface is dominated by the thermal conduction of the substrate [13].

Here we observe conductance quantization (Fig. 2(a)) in the $V_g$ – dependent conductance measurement at $T = 10$ K with $V = 1$ mV. The observations are consistent with the electrical conductance of a spin-degenerate quasi-1D system that can be described in Landauer formalism as

$$G = \frac{2e^2}{h} \sum_{n=1}^{\infty} \int T_n(E) \frac{-\partial f}{\partial E} dE, \qquad (3)$$

where $T_n(E)$ is the electron transmission probability through the $n^{th}$ 1D subband. [11,14] In the ballistic limit, the conductance is an integer of $2e^2/h$. Moreover, when sweeping $V$ and $V_g$, the short devices ($L \approx 200$ nm) exhibit a diamond shaped area with roughly constant differential conductance $g = dI/dV$ equal to the quantized conductance values (Fig. 2(b)), which is a feature of ballistic quasi-1D transport. [15,16] Outside the diamond, the number of occupied subbands is different at the two terminals, each with Fermi level $E_{F,S}$ and $E_{F,D}$, respectively, and $g$ deviates from the quantized values. [15,16] The top and bottom tips of the diamond shaped region correspond to when $E_{F,S}$ and $E_{F,D}$ align with two different subband edges (Fig. 2(c)), based on which an approximately 18 meV spacing between the 1st and 2nd subband can be extracted from Fig. 2(b). This value agrees with the numerical solution of the Schrödinger equation for a 60 nm diameter hexagonal hard-wall confinement with electron effective mass 0.026 $m_e$ [17]. The calculated radial probability density profiles of the 1st and the degenerate 2nd / 3rd subband and the corresponding radial confinement energies $E_1 = 12$ meV and $E_{2,3} = 30$ meV are shown in Fig. 2(d). In reality, the electrostatic potential in the presence of the source, drain, and gate contacts will break the rotational symmetry, and the 2nd / 3rd quantum state will no longer be degenerate. As shown in Fig. 2(a), the height of the 2nd conductance step is approximately equal to the 1st step.

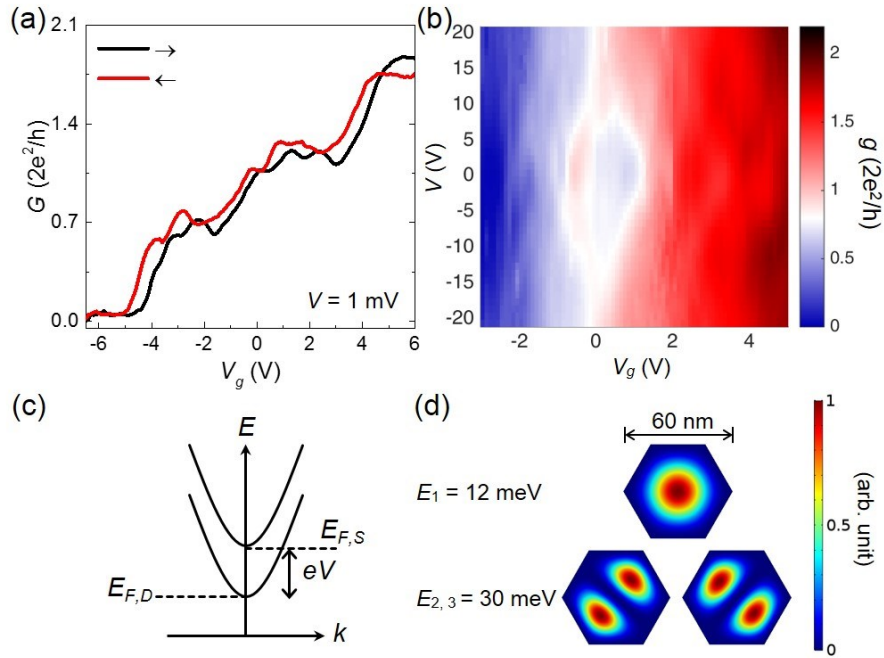

**Figure 2.** (a) $G$ ($V_g$) measured with increasing ($\rightarrow$) and decreasing ($\leftarrow$) $V_g$, and with $V = 1$ mV for a device with $L = 180$ nm. (b) Differential conductance $g = dI/dV$ measured as a function of $V_g$ and $V$ for a device with $L = 275$ nm. (c) A sketch of 1D subband dispersion relations marked with Fermi levels at the two contacts $E_{F,S}$ and $E_{F,D}$ aligning with two neighboring subband edges. (d) Calculated radial probability density profiles and radial confinement energies of the three lowest subbands under 60 nm diameter hexagonal confinement with 0.026 $m_e$ electron effect mass.

Then, in order to accurately extract $S^2G$, we measure $G$ and $V_{th} = S\Delta T$ (Fig. 3(a, b)) simultaneously from the slope and voltage offset at the open circuit condition ($I = 0$) of the $I - V$ curves [18] (Fig. 3(c)). In connection to the conductance quantization, $S$ shows oscillations that are characteristic of 1D subband transport. $S$ can be described in the Landauer formalism as [11]

$$S = G^{-1} \frac{2e}{hT} \sum_{n=1}^{\infty} \int T_n(E)(E - E_F) \frac{\partial f}{\partial E} dE. \tag{4}$$

The magnitude and sign of $S$ depend on the balance between electron transport above and below $E_F$. Inferring from Eq. (3), the onset of each $G$ step comes from the population of a new 1D subband, i.e. when $E_F$ is close to a subband edge (Fig. 3(a) inset (1)). Under this condition, $S$ is non-zero because within an energy range $\sim k_BT$ there are more transport channels for $E > E_F$. Conversely, the $G$ plateau occurs when $E_F$ is more than $\sim k_BT$ away from any subband edge (Fig. 3(a) inset (2)). In this case, $S$ will approach zero because within $\sim k_BT$ there are approximately as many transport above and below $E_F$. One exception being that when $E_F$ is decreased below the 1$^{st}$ subband edge, given that the valence band is far away, $S$ will increase continuously as there are no states below $E_F$. Overall, $S$ and the deduced $S^2G$ show a decaying oscillation as a function of $V_g$ (Fig. 3(d, e)). These features compare qualitatively well with theoretical and experimental studies of QPCs, [3,4] confirming the interpretation that the thermoelectric properties are dominated by quasi-1D transport. Yet the observed electronic transport is non-ballistic, and the scattering processes have a visible influence on the measured $G$ and $S$. For example, we observe a dip in the conductance before the onset of each conductance step and the concurrent sign change in $S$ (Fig. 3(a)), which resembles the theoretically predicted channel opening effect [19].

It is worth noting that in order to consider systems with different ballisticity, a generalized Landauer formalism [20,21] is used here, where the transmission $T_n(E)$ is an effective value that includes both elastic and inelastic scattering.

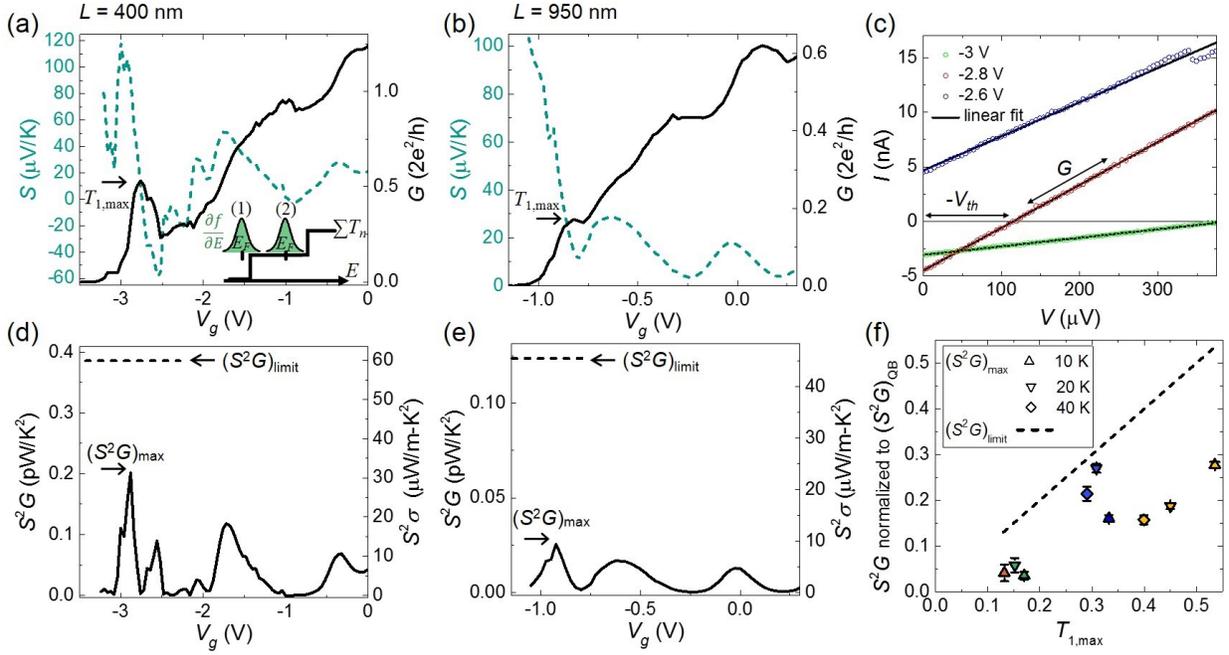

**Figure 3.** (a, b) $G$ ($V_g$) and $S$ ($V_g$) measured in $L$ = 400 and 950 nm devices. $T_{1,max}$ can be estimated from the height of the 1$^{st}$ conductance step. Inset: schematics of energy dependent electron transmission probability $\sum_n T_n(E)$ through a quasi-1D system, where $T_n$ ($E$) is assumed to be a constant. The $\frac{\partial f}{\partial E}$ distribution is plot for (1) $E_F$ within ~ $k_BT$ from the band edge, and (2) $E_F$ more than ~ $k_BT$ away from the band edges. (c) Linear fitting is used to extract the Seebeck voltage $V_{th}$ = $S\Delta T$ and conductance $G$ from the same $I$-$V$ curves. (d, e) $V_g$ - dependent power factor ($S^2G$ and $S^2\sigma$) plotted along with the theoretical limit $(S^2G)_{limit}$ (dashed lines). (f) Comparison of $(S^2G)_{max}$ with $(S^2G)_{limit}$ for devices with $L$ = 950 (green), 820 (orange), 310 (blue), and 400 nm (yellow) at $T$ = 10, 20, and 40 K.

As expected, the maximum $S^2G$, $(S^2G)_{max}$, is found near the depletion of the 1$^{st}$ subband (Fig. 3(d, e)). At temperatures 10 – 40 K, $k_BT$ is much smaller than the 1$^{st}$ - 2$^{nd}$ subband spacing (≈ 18 meV), therefore we attribute the measured $(S^2G)_{max}$ solely to electron transport through the 1$^{st}$ subband. The 1$^{st}$ conductance step heights (indicated by arrows in Fig. 3(a, b)) provide an estimation of the maximum electron transmission probability $T_{1,max}$ through the 1$^{st}$ subband, at least within the relevant energy range. Based on the found maximum, $(S^2G)_{limit}$ can be calculated (Eq. (1)). We find in Fig. 3(d-f) that $(S^2G)_{max}$ and $(S^2G)_{limit}$ are within the same order of magnitude and $(S^2G)_{max}$ shows an increasing trend with $T_{1,max}$, in agreement with $(S^2G)_{limit}$.

In the quasi-ballistic and diffusive transport regime, if we consider that the resistance due to different scattering process in the transport channel adds classically, the transmission probability of charge carriers follows [14,22,23]

$$T_n = \frac{\lambda_n}{L_c + \lambda_n}, \tag{5}$$

with electron mean free path $\lambda_n$ and channel length $L_c$. Consider that $L_c$ only deviates slightly from the device length $L$, by combining Eq. (2) and (5), $(S^2G)_{\text{limit}}$ can be modified to obtain the theoretical upper bound for $S^2\sigma$,

$$(S^2\sigma)_{\text{limit}} = (S^2G)_{\text{limit}}(S^2\sigma)_{\text{limit}} = (S^2G)_{\text{limit}} \times \frac{L}{A} < (S^2G)_{QB} \times \frac{\lambda_{1,\max}}{A}. \tag{6}$$

This expression directly connects the limit of $S^2\sigma$ in a 1D NW to the quantum limit of thermoelectric power production. $(S^2\sigma)_{\text{limit}}$, as opposed to $(S^2G)_{\text{limit}}$, can be used to compare with materials across different dimensions.

From the observed conductance step heights (Fig. 2(a), 3(a, b), 4(a)), we find $T_{1,\max} = 0.9 - 0.07$ for devices with $L = 180$ - $1240$ nm. By setting $L_c = L - 2\delta$, where $\delta$ is a fitting parameter that accounts for the downward band bending near the source and drain contact caused by the metal-semiconductor work function difference and sulfur passivation penetration and possible imperfect semiconductor-metal contact, we obtain $\overline{\lambda_{1,max}} = 232 \pm 81$ nm and $\bar{\delta} = 79 \pm 101$ nm by fitting the experimental values with Eq. (5) (Fig. 4(b)). Based on Eq. (6), we extract $(S^2\sigma)_{\text{limit}} \approx 72$ µW/m-K$^2$ for the ensemble of NWs measured in this study, which is consistent with the measured values $(S^2\sigma)_{\max} = 12 - 43$ µW/m-K$^2$. The value is relatively low compared to power factors of several mW/m-K$^2$ found in bulk materials with large charge effective mass. [24,25] Eq. (6), which highlights that small NW cross-section and long electron mean free path are needed to achieve large $S^2\sigma$, provides an explicit guide to understand what $S^2\sigma$ can be achieved with existing NWs. For example, InAs NWs with 28 nm × 40 nm cross-section area and 930 nm electron mean free path at 4 K were recently demonstrated. [26] According to Eq. (6), such NWs can be expected to have a competitive $S^2\sigma$ of around 0.8 mW/m-K$^2$ with single-subband transport at low temperatures, where high power factors are generally more difficult to achieve [7,27].

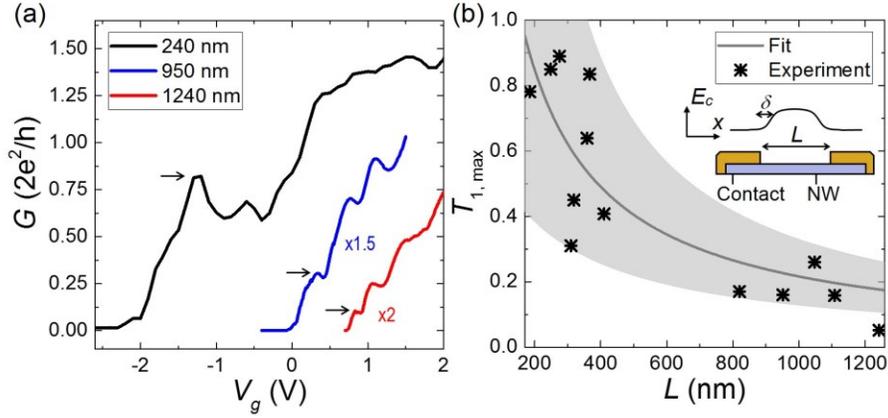

**Figure 4.** (a) $G(V_g)$ for three different devices with $L = 240$, 950, and 1240 nm. For devices with $L = 950$ and 1240 nm $G$ is scaled by 1.5 and 2.5 times, respectively. The arrows indicate the 1$^{st}$ conductance step height, which is used to extract $T_{1,max}$. (b) Extracted $T_{1,max}$ from devices with various $L$ (black asterisks), fitted with Eq. (5) (gray line). The gray area indicates the standard deviation of the fit. Inset: fitting parameter $\delta$ used to account for the downward band bending near the metal contacts.

In conclusion, we introduced and experimentally tested a theoretical limit for the power factors of non-ballistic 1D NWs. First, we showed that the quantum bound of thermoelectric power production leads to a stringent limit on the power factor, $S^2G$, of a non-ballistic 1D electronic system. Experimental observation of conductance quantization and Seebeck coefficient oscillation then allowed us to identify 1D electronic transport and extract the maximum $S^2G$ of the 1$^{st}$ subband, which conformed to the proposed limit. However, for practical applications, the thermoelectric power density is often of interest. Therefore, we also established the limit on the power factor, $S^2\sigma$, of an effective medium made of closely packed nanowires stretching into the diffusive transport regime. This limit provides an explicit guide on the optimal $S^2\sigma$ that can be achieved in realistic nanowire structures. These findings are helpful for quantitative predictions and to better inform and guide future efforts to improve the thermoelectric performance of 1D nanowires.

The authors thank Luna Namazi for providing the nanowires, and Martin Leijnse for helpful discussion, and acknowledge financial support from the People Programme (Marie Curie Actions) of the European Union's Seventh Framework Programme (FP7-People-2013-ITN) under REA grant agreement No 608153, PhD4Energy, from Marie Sklodowska Curie Actions Cofund, Project INCA 600398, by the Swedish Energy Agency (project P38331-1), by the Knut and Alice Wallenberg Foundation (project 2016.0089), and by NanoLund.

[1] L. D. Hicks and M. S. Dresselhaus, Phys. Rev. B **47**, 16631 (1993).